\documentclass[aps,preprint,showpacs,superscriptaddress]{revtex4}
\usepackage{amssymb}

\usepackage{graphicx}

\begin{document}
\title{  Decoy-state quantum key distribution with both source errors and statistical fluctuations}
\author{Xiang-Bin Wang}
\affiliation{Department of Physics, Tsinghua University, Beijing
100084, China} \affiliation{  Tsinghua National Laboratory for
Information Science and Technology, Beijing 100084, China}
\author{Lin Yang }
\affiliation{Department of Physics, Tsinghua University, Beijing
100084, China} \affiliation{Key Laboratory of Cryptologic Technology
and Information Security, Ministry of Education, Shandong
University, Jinan, China}
\author{Cheng-Zhi Peng}
\affiliation{Department of Physics, Tsinghua University, Beijing
100084, China}\affiliation{Hefei National Laboratory for Physical
Sciences at Microscale and Department of Modern Physics, University
of Science and Technology of China, Hefei, Anhui 230026, China}
\author{Jian-Wei Pan}
\affiliation{Department of Physics, Tsinghua University, Beijing
100084, China} \affiliation{Hefei National Laboratory for Physical
Sciences at Microscale and Department of Modern Physics, University
of Science and Technology of China, Hefei, Anhui 230026, China}
\affiliation{Physikalisches Institut, Universit\"at Heidelberg,
Philosophenweg 12, 69120 Heidelberg, Germany}

\begin{abstract}
 We
 show how to calculate the fraction of single photon counts of
  the 3-intensity decoy-state quantum cryptography
faithfully  with both statistical fluctuations and source errors.
Our results only rely on the bound values of a few parameters of the
states of pulses.
\end{abstract}


 \maketitle


\section{ Introduction}   Quantum
key distribution (QKD)\cite{BB84,GRTZ02,DLH06,ILM,gllp,scar,cai} has
now been extensively studied both theoretically and experimentally.
In practice, if one uses an imperfect single-photon source with a
lossy channel, the security is undermined by the
photon-number-splitting attack \cite{PNS,PNS1}. Fortunately,  this
can be managed by a number of methods\cite{ILM,rep, H03, wang05,
LMC05,HQph,pdc1,pdc2,haya,scran,kko,zei}. In particular, the so
called ILM-GLLP proof\cite{ILM,gllp} has shown that if we know the
upper bound of fraction of the multi-photon counts (or equivalently,
the lower bound of single-photon counts) among all raw bits, we
still have a way to distill the secure final key. Verifying such a
bound is strongly non-trivial. The so called decoy-state method
\cite{rep, H03, wang05, LMC05,HQph} is  to find out such bounds,
say, among all those clicks at Bob's side,   at least how many of
them are due to the single-photon pulses from Alice.


The main idea of the decoy-state method is to change intensities
randomly among different values in sending out each pulses.
Equivalently, we can regard pulses of different intensities as
pulses from different sources. For example, in a 3-intensity
protocol\cite{wang05}, Alice has three sources, source $Y_0$ (vacuum
source) which is supposed to produce vacuum only, source $Y$ (decoy
source) which is supposed to only produce state
\begin{equation}\label{decoyd}
\rho=\sum_{k=0}^J a_k|k\rangle\langle k|
\end{equation}
only, and source $Y'$ (signal source) which is supposed to only
produce state
\begin{equation}\label{signald}
\rho'=\sum_{k=0}^J a_k'|k\rangle\langle k|,
\end{equation}
 where $|k\rangle$ is the $k-$photon Fock state and $a_k\ge
0,\;a_k'\ge 0$ for all $k$, $\sum a_k=\sum a_k'=1$. Here $J$ can be
either finite or infinite. Given a coherent-state source or a
heralded single-photon source from the parametric down conversion,
$J=\infty$.

The fundamental formula in the decoy-state method as first proposed
by Hwang\cite{H03} is that the counting rates (yields) of pulses of
the same photon-number states  must be equal to each other even they
are from different source, i.e.
\begin{equation}\label{0011cc}
s_k=s_k'
\end{equation}
where $s_k$ and $s_k'$ are counting rate of $k-$photon pulses from
the decoy source and counting rate of $k-$photon pulses from the
signal source, respectively. This is because of the obvious fact
that those $k-$photon pulses from different sources are actually
{\em randomly} mixed therefore they are {\em random samples} of each
other, {\em if} sources $Y_0,Y,Y'$ are perfectly stable, i.e., they
always produce a state precisely as the assumed one. Given
Eq.(\ref{0011cc}), one can list simultaneous equations\cite{wang05}
for $s_k$ based Eqs.(\ref{decoyd},\ref{signald}) and then find out
the lower bound of $s_1$ ($s_1'$) which indicates the lower bound of
single photon counts of the decoy source and signal source.

Given this, one can calculate the final key rate of each source, by
the well known result of
 ILM-GLLP\cite{ILM,gllp}.
 For example, the asymptotic final
 key rate for the signal source can be calculated by
\begin{equation}\label{ilm}
R_s=\Delta_1'[1-H(t_1)]-H(t)
\end{equation}
where
 $t_1$, $t$ are the QBER for single-photon counts of the signal source and the QBER for
all counts of the signal pulses from the signal source. $\Delta_1'$
is the lower bound of the fraction of counts due to single photon
pulses from the signal source. Moreover, the non-asymptotic
unconditional secure key rate in a finite-size QKD is extensively
studied recently, first in Ref.\cite{ILM}, and then further studied
in \cite{scar,cai}.

Recently, a number of QKD experiments using the decoy-state method
have been done \cite{ron,peng,sch,yuan,tanaka}.  However, the story
is not completed here because the existing decoy-state theory does
not entirely cover the real experimental conditions in practice. No
source in real-life can be perfectly  stable. One important problem
is the effect of source errors.  Suppose in a protocol Alice sends
out $M$ pulses one by one. The actually produced state at any time
$i$ can differ from the assumed state that she {\em wants}. We call
this discrepancy {\em source error}. Say, at any time $i$, Alice
{\em wants} to prepare a state
\begin{equation}
\rho_i = \sum_{k=0}^J f_k|k\rangle\langle k|
\end{equation}
for Bob, but the source actually prepares a slightly different state
\begin{equation}
\tilde\rho_i = \sum_{k=0}^J f_{ki}|k\rangle\langle k|
\end{equation}
in the Fock space. We shall call such $f_{ki}-f_k$ source error, or
state error, or parameter fluctuation. Most generally, the error,
$f_{ki}-f_k$ is {\em not} random for different $i$. In practice, one
of the major cause of the source error is the  intensity
fluctuation.  Therefore we often use the term {\em intensity
fluctuation} for the source errors. But our result here is not
limited to the intensity fluctuation only. Here we consider  a more
general case that the parameters in the state fluctuate. Therefore
we shall still use the term {\em source error} in stead of intensity
fluctuation.

If the source error is random,
    we can simply assume a perfectly stable source always emitting the averaged-state\cite{wang07,HQph}
    of all pulses from a certain source. As shown in Ref.\cite{wang07},
    we have to take a feed-forward control to each individual pulses in order to guarantee the randomness of source errors.
    If we use coherent light only, and if we have a stable two-value attenuator,
     we can use the method in Ref.\cite{wangapl} to manage
    any intensity fluctuations.

If we don't assume any conditions above, the issue need to be
studied more carefully. A very tricky point here is that the
elementary assumption of Eq.(\ref{0011cc}) for the decoy-state
method with stable sources is in general incorrect, if the state
errors are {\em not} random.
    As we have
shown explicitly\cite{rep,tdecoy}, there are cases that Eve. may
know the source errors and she can violate Eq.(\ref{0011cc}) by
producing an instantaneous transmittance channel dependent on the
errors. Therefore, we must seek a  solution to the new problem.

 Very
recently\cite{tdecoy}, a general asymptotic theory for decoy-state
QKD with source errors is presented. By the method\cite{tdecoy},  we
don't have to change anything in the existing set-up. The only thing
we need is the bound values of a few parameters in the states from
each source. However, that work has only presented the asymptotic
result i.e., the effect of {\em classical statistical fluctuation}
is not considered. Besides Ref.\cite{tdecoy}, the problem is also
studied studied\cite{yi} asymptotically from another viewpoint for
the Plug-and-Play protocol. There\cite{yi}, Alice receives a pulse,
attenuates it and sends it out for Bob. Eve controls both the
incident pulse and the outcome pulse. Alice's transmittance $\gamma$
in doing attenuation can be either $\gamma^D$ or $\gamma^S$,
depending on whether she wants to prepare a decoy pulse or a signal
pulse. Similar to Ref.\cite{wangapl}, Ref.\cite{yi} also assumes
that Alice can do the two-value attenuation exactly.    The
fundamental formula assumed  there \cite{yi}is
$$
Y_{m,n}^D=Y_{m,n}^S
$$
where $Y^D_{m,n}$($Y_{m,n}^S$) is the counting rate of a
decoy(signal) pulse which contains $m$ photons when flying into
Alice lab and contains $n$ photons when flying away from Alice's
lab, after Alice's attenuation. As was pointed out by the authors of
Ref.\cite{yi}, the effects of the internal fluctuation of Alice's
lab was not considered there and also the validity of elementary
equation $Y_{m,n}^D=Y_{m,n}^S$ is not studied\cite{comment}.
Actually, as shown in the appendix of this work, the elementary
equation does not always hold if there are fluctuations to Alice's
attenuation.

 Here in this paper, we study the effects of {\em both} source errors {\em and}
 classical statistical fluctuations. The existing works\cite{tdecoy,yi}
 have not included the effect of statistical fluctuations in calculating the final
 key rate, though the effect can be included in principle.
   Our earlier work
 assumes zero error for vacuum source $Y_0$, here we shall assume errors  in this source, besides
 errors in source $Y$ and source $Y'$.
 Similar to the results in \cite{tdecoy}, the results presented
 here only need the bound values of a few parameters in the states of different
 sources. In particular, in deriving the fraction of single-photon counts
 we don't use any unproven assumption and we don't need worry about the
internal fluctuation of any Alice's device, since our method {\em
only needs the bound values of a few parameters in the source
state}. We don't need to presume any specific distribution for our
 states.
 In our study in this work, classical statistical fluctuation is considered by estimating the observed values
  from the asymptotic values with a fixed standard deviation. We shall start with a protocol with one-way quantum
communication only. But as we point out in the end of this paper,
our method obviously also applies to a Plug-and-Play protocol.

 This paper is arranged as the following.
After the introduction above, we present  our method with some
important mathematical relations section II. We then present our
main results in section III: the asymptotic and non-asymptotic
formula for the fraction of single photon counts of the 3-intensity
protocol with errors in states of all 3 sources. Some remarks on the
source errors are addressed in section IV. The paper is ended with a
concluding remark.


\section{Our method}\label{goal}
We  assume that Alice sends $M$ pulses to Bob one by one in the
protocol. Each pulse sent out by Alice is randomly chosen from one
of the 3 sources $Y_0,Y,Y'$ with constant probability $p_0,p,p'$,
respectively. Each source has errors. Clearly,  $pM$, $p'M$ and
$p_0M$ are just the number of the decoy pulses, the number of the
signal pulses, and the number of pulses from the vacuum source,
respectively. We assume we know the bounds of parameters in the
states of each sources.

Our {  goal} is to find out the lower bound of the fraction of
counts caused by those single-photon pulses for both the signal
source and the decoy source. The following quantities  are directly
observed in the protocol therefore we regard them as known
parameters: $N_d$, the number of counts caused by the decoy source,
$N_s$, the number of counts caused by the signal source, and $N_0$,
the number of counts caused by the vacuum source, $Y_0$.
  For our goal, we
only need to formulate the number of counts caused by those
single-photon pulses from each sources in terms of the known
quantities $N_0,N_d,N_s$ and $p_0,p,p'$ and the bound values of
those parameters of states in each sources.



%
\subsection{Virtual protocol}
For clarity, we
 first
consider a virtual protocol;\\ Suppose Alice will send $M$ pulses to
Bob in the whole protocol. At any time $i$ ($i\in
 [1,M]$), each source produces a pulse. The states of the pulses from
sources $Y_0,Y\;,Y'\;$ are
\begin{equation}
\rho_{0i}= \sum_{k=0}^J b_{ki} |k\rangle\langle k|,\label{rho0i}
\end{equation}
\begin{equation}\label{rhoi}\rho_i=\sum_{k=0}^J a_{ki}|k\rangle\langle k|\; ;\;{\rm and}\end{equation}
\begin{equation}\label{rhoip} \rho_i'=\sum_{k=0}^J a_{ki}'|k\rangle\langle k|.\end{equation}
Here $b_{0i}$ is a bit smaller than 1, and we assume $b_0^L$, lower
bound of all $b_{0i}$ is known in the protocol;  $\rho_i$ and
$\rho_i'$ are a bit different from $\rho$ and $\rho'$ of
Eq.(\ref{decoyd}, \ref{signald}), which are the assumed states in
the perfect protocol where there is no source error. At any time
$i$, only one pulse is selected and sent out for Bob. The
probability of selecting the $i$th pulse  source $Y_0,Y$ or $Y'$  is
constantly $p_0,\;p$, and $p'$ for  any $i$.  The un-selected two
pulses at each time will be discarded. After Bob has completed all
measurements to the incident pulses, Alice checks the record about
which pulse is selected at each time, i.e., which time has used
which source.  Obviously, Alice can decide which source to be used
at
each time in the very beginning. This is just then the real protocol
of the decoy-state method.

As shown below, based on this virtual protocol, we can formulate the
number of counts from each source and therefore find the lower bound
of the number of single-photon counts. The result also holds for the
real protocol where Alice decides to use which sources at the $i$th
time in the very beginning.

\subsection{Some definitions}
{\em Definition 1}. In the protocol, Alice sends Bob $M$ pulses, one
by one. In response to Alice, Bob observes his detector for $M$
times. As Bob's {\em i}th observed result, Bob's detector can either
click or not click. If the detector clicks in Bob's {\em i}th
observation, then we say that ``the {\em i}th pulse from Alice has
caused a count". We disregard how the {\em i}th pulse may change
after it is sent out. When we say that Alice's {\em i}th pulse has
caused a count we only need Bob's detector clicks in Bob's {\em i}th
observation.

Given the source state in Eqs.(\ref{rhoi},\ref{rhoip}), any $i$th
pulse sent out by Alice must be in a photon-number state. To anyone
outside Alice's lab, it looks as if that Alice only sends a photon
number state at each single-shot: sometimes it's vacuum, sometimes
it's a single-photon pulse, sometimes it is a $k-$photon pulse, and
so on. We shall make use of this fact that any individual pulse is
in one Fock state.
\\
{\em Definition 2}. Set $C$ and $c_k$:  Set $C$ contains any pulse
that has caused a count; set $c_k$ contains any $k-$photon pulse
that has caused a count. Mathematically speaking, the sufficient and
necessary condition for $i\in C$ is that  the {\em i}th pulse has
caused a count.  The sufficient and necessary condition for $i\in
c_k$ is that  the {\em i}th pulse contains $k$ photons and it has
caused a count. For instance, if the photon number states of the
first 10 pulses from Alice are
$|0\rangle,\;|0\rangle,\;|1\rangle,\;|2\rangle,\;|0\rangle,\;
|1\rangle,\;|3\rangle,\;|2\rangle,\;|1\rangle,\;|0\rangle,\;$  and
the pulses of $i=2,\; 3,\; 5,\; 6,\; 9,\; 10$ each has caused a
count at Bob's side, then we have
\begin{equation}
C=\{i|i=2,\;3,\;5,\;6,\;9,\;10,\cdots\};\;c_0=\{i|i=2,5,10,\cdots\};
\; c_1=\{i|i=3,6,9,\cdots\}.
\end{equation}
Clearly, $C=c_0\cup c_1\cup c_2\cdots$, every pulse in set $C$ has
caused a count.
\\{\em Definition 3}. We use superscripts $U,L$ for the upper bound and lower bound of a certain
parameter. In particular, given any $k\ge 0$ in
Eqs.(\ref{rho0i},\ref{rhoi}, \ref{rhoip}), we denote
$x_{k}^L,x_{k}^U$ for the minimum value and maximum value of
$\{x_{ki}|\;i\in c_k\}$; and $x=b,a,a'$.  We assume these bound
values are known in the protocol.
\subsection{Some important relations and facts}
If the $i$th pulse is an element of $c_k$, the probability that it
is from $Y_0$, $Y$ or $Y'$ is
\begin{eqnarray}\label{prob}
\mathcal
P_{vi|k}=\frac{p_0b_{ki}}{p_0b_{ki}+pa_{ki}+p'a_{ki}'},\nonumber\\
\mathcal
P_{di|k}=\frac{pa_{ki}}{p_0b_{ki}+pa_{ki}+p'a_{ki}'},\nonumber\\
\mathcal P_{si|k}=\frac{p'a_{ki}'}{p_0 b_{ki}+pa_{ki}+p'a_{ki}'}.
\end{eqnarray}

 We
want to formulate the numbers of $k$-photon counts  caused by each
sources. Given the definition of the set $c_k$, this is equivalent
to ask how many of pules in set $c_k$ come from each sources.
 If the $i$th pulse contain $k$ photons, it can come from any
of  the 3 sources,  $Y_0$, $Y$ or $Y'$. According to
Eqs.(\ref{rho0i}, \ref{rhoi}, \ref{rhoip}), if the {\em i}th pulse
contains $k$-photons, the probability that it comes from source
$Y_0$ is
$$\mathcal P_{vi|k}=\frac{b_{ki}p_0}{p_0b_{ki}+pa_{ki}+p'a_{ki}'}.$$
Or, equivalently,
$$ n_{k0}=\sum_{i\in c_k}\mathcal P_{vi|k}.$$
Given a finite number of pulses and counts, this equation should be
replaced by the expectation-value equation as
\begin{equation}
\langle N_{0}\rangle = \sum _{k=0} ^J \sum_{i\in c_k}\mathcal
P_{vi|k}
\end{equation}
 This is  the expected number of counts caused by source $Y_0$, since
 every pulse in $\{c_k\}$ has caused a count.
 Therefore we can formulate the expected value of the
number of counts caused by source $Y_0$ by
\begin{equation}\label{vcct}
\langle N_0\rangle  = \sum_{k=0}^J\sum_{i\in c_k}
b_{ki}p_0d_{ki}=\langle N_{0}^* \rangle+\sum_{k=1}^J\sum_{i\in c_k}
b_{ki}p_0d_{ki}
\end{equation}
Here
\begin{equation}\label{ddki}
 d_{ki}=\frac{1}{p_0b_{ki}+pa_{ki}+p'a_{ki}'}.
\end{equation}
$N_0$ is the number of counts due to pulses from source $Y_0$,
$N_0^*=p_0\sum_{i\in c_0}b_{0i}d_{0i}$ is the number of counts due
to those vacuum pulses from source $Y_0$. Similarly, if the {\em
i}th pulse contains $k$ photons, it has a probability $ \mathcal
P_{di|k}=\frac{pa_{ki}}{p_0b_{ki}+pa_{ki}+p'a_{ki}'}$ to be from the
decoy source, and a probability of $ \mathcal
P_{si|k}=\frac{p'a_{ki}'}{p_0b_{ki}+pa_{ki}+p'a_{ki}'}$ to be from
the signal source. Therefore we have
\begin{equation}\label{s0d2}
\langle n_{kd} \rangle =  \sum_{i\in c_k} \mathcal
P_{di|k}=\sum_{i\in c_k}pa_{ki}d_{ki},
\end{equation}
for the expected number of counts caused by those $k$-photon pulses
from the decoy source, and
\begin{equation}\label{s0s2}
\langle n_{ks}'\rangle  =\sum_{i\in c_k} \mathcal P_{si|k}=
\sum_{i\in c_k}p'a_{ki}'d_{ki} ,
\end{equation}
for the expected number of counts caused by those $k$-photon pulses
from the signal source. Therefore, besides Eq.(\ref{vcct}) we also
have the following 2 equations for   $\langle N_d\rangle$ and
$\langle N_s\rangle$ as the expected values of the number of counts
due to the decoy pulses and signal pulses:
\begin{equation}\label{popd}
\langle N_d\rangle=\sum_{k=0}^J\sum_{i\in c_k}\mathcal
P_{di|k}=p\sum_{k=0}^J\sum_{i\in c_k}a_{ki}d_{ki}
\end{equation}
\begin{equation}\label{pops}
\langle N_s\rangle=\sum_{k=0}^J\sum_{i\in c_k}\mathcal
P_{si|k}=\sum_{k=0}^J\sum_{i\in c_k}a_{ki}'d_{ki}
\end{equation}
 Here we have used
Eqs.(\ref{s0d2}, \ref{s0s2}).

For simplicity, we define
 \begin{equation}\label{Dki}
 D_k=\sum_{i\in c_k} d_{ki}=\sum_{i\in
 c_k}\frac{1}{p_0b_{ki}+pa_{ki}+p'a_{ki}'}.
 \end{equation}
Recall Eqs.(\ref{s0s2},\ref{s0d2}), $\langle n_{0s}'\rangle,\;
\langle n_{0d}\rangle$ are the expected number of pulses from the
decoy source and signal source in set $c_0$.

Based on the formulas and definitions above, we find the following
important facts:
\\{\em Fact 1:}
\begin{equation}
 pa_0^UD_0 \ge \langle n_{0d}\rangle \ge pa_0^LD_0;\;\;
 p'a_0'^UD_0 \ge \langle n_{0s}'\rangle \ge p'a_0'^LD_0.
\end{equation}
This is directly deduced from Eqs.(\ref{s0d2},\ref{s0s2}). \\
{\em Fact 2:}
\begin{equation}
\frac{\langle N_{0} \rangle}{b_0^Lp_0} \ge D_0\ge
\frac{a_1^L}{p_0\left[ a_1^L-a_0^L(1-b_0^L)\right]}\left(\langle
N_0\rangle -\frac{p_0(1-b_0^L) }{pa_1^L}\langle N_{d} \rangle
\right).
\end{equation}
  {\em Proof}: Obviously,
\begin{equation}\label{n00}
D_0 \le \frac{\langle N_{0} \rangle}{b_0^Lp_0}.
\end{equation}
On the other hand, since any $b_{0i}\le 1$,
\begin{equation}\label{med1}
p_0 D_0 \ge \langle N_0\rangle - p_0\sum_{k=1}^J\sum_{i\in
c_k}b_{ki}d_{ki} \ge \langle N_0\rangle -
p_0\sum_{k=1}^Jb_{k}^UD_{k}
\end{equation}
To lower bound $D_0$, we only need upper bound
$p_0\sum_{k=1}^Jb_{k}^UD_{k}$.  By Eq.(\ref{popd}) we know
\begin{equation}
p\sum_{k=1}^J a_k^LD_k\le \langle N_d\rangle -\langle n_{0d}\rangle
\end{equation}
which is equivalent to say
\begin{equation}\label{med2}
p_0\frac{b_1^U}{a_1^L}\sum_{k=1}^J a_k^LD_k\le
\frac{p_0b_1^U}{pa_1^L}\left(\langle N_d\rangle -\langle
n_{0d}\rangle\right)
\end{equation}
We assume
\begin{equation}\label{ace00}
\frac{a_k^L}{b_k^U}\ge \frac{a_1^L}{b_1^U}.
\end{equation}
This condition can obviously hold if  each pulses of source $Y_0$ is
in an extremely weak coherent state. These conditions mean
\begin{equation}
p_0\sum_{k=1}^Jb_{k}^UD_{k} \le
\frac{p_0b_1^U}{pa_1^L}\left(p\sum_{k=1}^J a_k^LD_k\right)
\end{equation}
Therefore, based on Eq.(\ref{med1},\ref{med2}) we have
\begin{equation}\label{med3}
p_0 D_0  \ge \langle N_0\rangle - \frac{p_0b_1^U}{pa_1^L}\left(
\langle N_d\rangle -\langle n_{0d}\rangle\right)\ge \langle
N_0\rangle - \frac{p_0(1-b_0^L)}{pa_1^L}\left( \langle N_d\rangle
-\langle n_{0d}\rangle\right)
\end{equation}
Combining {\em fact 1} with Eq.(\ref{med3}), we obtain the following
important formula
\begin{equation}
 D_0 \ge \frac{a_1^L}{p_0\left[ a_1^L-a_0^L(1-b_0^L)\right]}\left(\langle
N_0\rangle -\frac{p_0(1-b_0^L) }{pa_1^L}\langle N_{d} \rangle
\right).
\end{equation}
This completes the proof of fact 2. \\
{\em Fact 3}:
\begin{eqnarray}
D_1 \ge \mathcal D_1^L=\frac{a_2'^L\langle N_d\rangle/p-a_2^U\langle
N_s\rangle/p'-\left(a_2'^L a_0^U-a_2^Ua_0'^L\right) D_0
}{a_1^Ua_2'^L-a_1'^La_2^U} .\label{D11}
\end{eqnarray}
{\em Proof}: The startpoint of our proof is Eqs.(\ref{popd},
\ref{pops}) which can be rewritten into
\begin{eqnarray}\label{d0d}
\langle N_d\rangle=
\langle n_{0d}\rangle+pa_{1}^UD_1 +p\Lambda -\xi_1\\
\langle N_s\rangle=\langle n_{0s}'\rangle+ p'a_{1}'^LD_1 +p'\Lambda'
+\xi_2 \label{s0s}\end{eqnarray} where
\begin{equation}
\Lambda=\sum_{k=2}^J a_{k}^U\sum_{i\in c_k} d_{ki};\;
\Lambda'=\sum_{k=2}^J a_{k}'^L\sum_{i\in c_k}d_{ki},
\end{equation}
and
  \begin{eqnarray}\xi_1=p\left[a_{1}^UD_1 +\Lambda-
\left(\sum_{i\in c_1} a_{1i}d_{1i} +\sum_{k=2}^J\sum_{i\in c_k} a_{ki}d_{ki}\right)\right]\ge 0\nonumber\\
\xi_2=p'\left[\sum_{i\in c_1} a_{1i}'d_{1i} +\sum_{k=2}^J \sum_{i\in
c_k}a_{ki}'d_{ki}- \left( a_{1}'^L D_1 + \Lambda'\right)\right]\ge
0\nonumber
\end{eqnarray}
According to the definition of $\Lambda$ and $\Lambda'$, we also
have
\begin{equation}\label{lambda}
\Lambda'=\frac{a_2'^L}{a_2^U}\Lambda + \xi_3
\end{equation}
and \begin{equation}\xi_3= \Lambda'-\frac{a_2'^L}{a_2^U}\Lambda
\end{equation} Further, we assume the important condition
\begin{equation}\label{ace}
\frac{a_k'^L}{a_k^U}\ge  \frac{a_2'^L}{a_2^U}\ge
\frac{a_1'^L}{a_1^U},\;{\rm for\; all}\;\; k\ge 2.
\end{equation}
The first inequality above leads to
\begin{equation}
\xi_3\ge 0
\end{equation}
as one may easily prove. With Eq.(\ref{lambda}),  Eq.(\ref{s0s}) is
converted to
\begin{eqnarray}
\langle N_s\rangle=\langle n_{0s}'\rangle+ p'a_{1}'^LD_1
+p'\frac{a_2'^L}{a_2^U}\Lambda +\xi_2+p'\xi_3
\label{s0s1}\end{eqnarray} Given the Eqs.(\ref{d0d}, \ref{s0s1}),
 we can formulate $D_1$:
\begin{equation}
D_1=\frac{a_2'^L\langle N_d\rangle/p-a_2^U\langle
N_s\rangle/p'-a_2'^L\langle n_{0d}\rangle/p+a_2^U\langle
n_{0s}'\rangle/p'+a_2'^L\xi_1/p+a_2^U(\xi_2+p'\xi_3)/p'}{a_1^Ua_2'^L-a_1'^La_2^U}.
\end{equation}
Since $\xi_1,\xi_2,$ and $\xi_3$ are all non-negative, and
$a_1^Ua_2'^L-a_1'^La_2^U\ge 0$ by the second inequality of
Eq.(\ref{ace}), we now have
\begin{eqnarray}
D_1=\sum_{i\in c_1} d_{1i} \ge  \frac{a_2'^L\langle
N_d\rangle/p-a_2^U\langle N_s\rangle/p'-a_2'^L\langle
n_{0d}\rangle/p+a_2^U\langle
n_{0s}'\rangle/p'}{a_1^Ua_2'^L-a_1'^La_2^U}\nonumber\\
\ge  \frac{a_2'^L\langle N_d\rangle/p-a_2^U\langle
N_s\rangle/p'-\left(a_2'^La_0^U -
a_2^Ua_0'^L\right)D_0}{a_1^Ua_2'^L-a_1'^La_2^U}=\mathcal D_1^L .
\end{eqnarray}
Here we have used {\em Fact 1} for the bound values of $\langle
n_{0d}\rangle,\;\langle  n_{0s}'\rangle$. To minimize $D_1$, we have
replaced $\langle n_{0d}\rangle$ by its upper bound and $\langle
n_{0s}'\rangle$ by its lower bound as given in {\em Fact 1}. This
completes the proof of {\em fact 3}.
\section{ Main results}
\subsection{Asymptotic result}
Given {\em Fact 3}, the minimum value of the number of counts caused
by single-photon pulses from the signal-source (or the decoy-source)
is simply
\begin{equation}\label{n1ds} \langle n_{1s}'\rangle^L = p'a_{1}'^LD_1\le \langle n_{1s}'\rangle,\;
{\rm (or\;} \langle n_{1d}\rangle^L =pa_1^LD_1\le \langle
n_{1d}\rangle {\rm )}.
\end{equation}
Therefore, we can now bound the fraction of single photon counts
among all counts caused by the signal source
\begin{equation}\label{main0}
\Delta_1'\ge \frac{p'a_1'^L\mathcal D_1^L}{\langle N_s\rangle}=
\frac{a_1'^L(a_2'^L\langle N_d\rangle p'/p-a_2^U\langle
N_s\rangle-p'\left(a_2'^L a_0^U-a_2^Ua_0'^L\right) D_0}{\langle
N_s\rangle(a_1^Ua_2'^L-a_1'^La_2^U)} .
\end{equation}
Here the range of $D_0$ is given by {\em fact 2} in the earlier
subsection. We don't have to replace $D_0$ by its largest possible
value to obtain the smallest possible $\Delta_1'$ at this moment.
Instead, we shall do numerical calculation by Eq.(\ref{ilm}) with
all possible values of $D_0$ in the range given by {\em fact 2} and
find the worst case result directly to the key rate.
 Define $S'=\frac{\langle N_s\rangle}{p'M}$ as the counting rate of
the signal source, $S=\frac{\langle N_d\rangle}{pM}$ as the counting
rate of the decoy source, $S_0=\frac{\langle N_0\rangle}{p_0M}$ as
the counting rate of source $Y_0$, and $M$ is the total number of
pulses as defined earlier, we can write the right-hand-side of the
inequality in term of counting rates:
\begin{equation}\label{main1}
\Delta_1'\ge
\frac{a_1'^L\left[a_2'^LS-a_2^US'-(a_2'^La_0^U-a_2^Ua_0'^L)S_0/b_0^L\right]}
{S'\left(a_1^Ua_2'^L-a_1'^La_2^U\right)}
\end{equation}
 Similarly, we also have
\begin{equation}\label{main2}
\Delta_1\ge
\frac{a_1^L\left[a_2'^LS-a_2^US'-(a_2'^La_0^U-a_2^Ua_0'^L)S_0/b_0^L\right]}
{S\left(a_1^Ua_2'^L-a_1'^La_2^U\right)}
\end{equation}
for the minimum value of fraction of single-photon counts for the
decoy source.

Eqs. (\ref{main0}, \ref{main1}, \ref{main2}) and conditions of
Eq.(\ref{ace00},\ref{ace}) are the asymptotic results of this work.


\subsection{Non-asymptotic results}
Strictly speaking, Eqs.(\ref{n1ds}, \ref{D11}) cannot be used in a
real experiment where the number of pulses are always finite.
Therefore, in any real experiment we have to consider the effect of
statistical fluctuation besides the effect of source errors.

First, $\langle n_{1s}'\rangle^L = p'a_1'^L D_1\ge  p'a_1'^L
\mathcal D_1^L$ in the formulas is the  lower bound of the
 expected number of single photon counts of the signal
source, but we actually need the lower bound of $n_{1s}'$, the
(deduced)  observed  value (i.e., the true value) of the number
single photon counts of the signal source in a certain experiment.
Of course, given the expectation value, $\langle n_{1s}'\rangle^L $,
one can deduce the lower bound $n_{1s}'^L$ of the experimental value
$n_{1s}'$ by classical statistics:
\begin{equation}
n_{1s}'\ge n_{1s}'^L=\langle n_{1s}'\rangle -\delta_1.
\end{equation}
To be sure that this bound is correct with a probability exponential
close to 1, we need set $\delta_1$ considerably large, e.g.,
\begin{equation}
\delta_1=10\sqrt{\langle n_{1s}'\rangle}.
\end{equation}
Here 10 standard deviation is chosen simply for convenience rather
than a exact cut off point.  As shown in Ref.\cite{wang05}, the
probability that the statistical fluctuation exceeds this range is
exponentially close to 0.

 Second, in the right hand side of Eq.(\ref{D11}), there are many
parameters of expectation values such as $\langle N_d\rangle,
\langle N_s\rangle, \langle n_{0d}\rangle,\langle n_{0s}'\rangle$,
which are {\em not}  observed values in one specific experiment. We
need to replace these parameters by the experimentally observed
quantities to make the formula useful. Again, this can be done by
using classical statistics. This is to say, we need consider the
largest possible difference between the expectation values and the
 observed values due to the statistic fluctuations.

For clarity, we shall use $\langle \zeta \rangle$ for the
statistical expectation value of observable $\hat\zeta$, and $\zeta$
for the true value or the observed value in the experiment. For
example, we shall use $\langle n_{1d}\rangle$ for the expectation
value of the number of counts at Bob's side due to  the
single-photon pulses from the decoy source at Alice's side. We shall
also use the superscript $U$ or $L$ to indicate the maximum value or
minimum value of certain variable.
  With the following condition,
\begin{equation}\label{norm1}
\mathcal P_{vi|k}+\mathcal P_{di|k} + \mathcal P_{si|k}=1, \; {\rm
if} \;i\in c_k
\end{equation}
for any $k$, we can now start to derive our non-asymptotic result.
By classical statistics we know that there exists a real number
$\delta_d$ satisfying
\begin{equation}
 N_d = \langle N_d \rangle +
\delta_d
\end{equation}
with a probability exponentially close to 1 if we set
\begin{equation}\label{range}
|\delta_d| \le \delta_d^U=10\sqrt {\langle N_d \rangle}\approx
10\sqrt { N_d }.
\end{equation}
This indicates that
\begin{equation}\label{ndr}
  \langle N_d\rangle = N_d -\delta_d
\end{equation}
with a probability exponentially close to 1 that $\delta_d$ is in
the range given by Eq.(\ref{range}). Also, there exists a real
number $\delta_0$ satisfying
\begin{equation}\label{b0}
 \langle N_0 \rangle = N_0 - \delta_0
\end{equation}
and $ |\delta_0| \le 10\sqrt{N_0}$.

In our problem, due to the  constraint given in Eq.(\ref{norm1}),
the fluctuations of each variables  are not independent. This can
help us not to overestimate the effects of the fluctuations too
much.
 Eq.(\ref{norm1}) immediately leads to
the following identity:
\begin{equation}\label{id}
\langle N_s\rangle +\langle N_d\rangle + \langle N_0\rangle=
N_s+N_d+N_0.
\end{equation}
This is to say, the total population in set $C$ is fixed, but there
could be fluctuation in the population distribution over sources
$Y_0,Y,Y'$. Eq.(\ref{id}), together with Eq.(\ref{ndr}) and
Eq.(\ref{b0}) leads to
\begin{equation}\label{nsr}
  \langle N_s\rangle = N_s - \delta_d - \delta_0.
\end{equation}
We use
\begin{equation} \label{tns}
n_{1s}'\ge \tilde n_{1s}'=\langle n_{1s}'\rangle^L -10\sqrt {\langle
n_{1s} \rangle^L}.
\end{equation}
This is  the inequality for the experimental lower bound of the
number of single-photon counts from decoy pulses, where
\begin{equation}\label{main22}
\langle n_{1s}'\rangle^L= p'a_1'^L \mathcal D_1^L\ge
\frac{p'a_1'^L\left[a_2'^L \langle N_d\rangle/p-a_2^U\langle
N_s\rangle/p'-\left(a_2'^L a_0^U-a_2^Ua_0'^L\right)\langle
D_0\rangle \right]}{a_1^Ua_2'^L-a_1'^La_2^U}
\end{equation}
and $\langle N_d\rangle,\;\;\langle N_s\rangle$ are defined by
Eqs.(\ref{ndr},\ref{nsr}). The range of $\langle D_0\rangle$ is
given by {\em Fact 2} where the range of  $\langle D_0\rangle$ is
given by Eq.(\ref{b0}). To have the lower bound value of $n_{1s}'$
in Eq.(\ref{tns}) , we only need put the smallest value of $\langle
N_d\rangle$ and largest possible values of $\langle N_s\rangle,
\langle D_0\rangle$ into Eq.(\ref{main22}). However, we don't have
use the largest possible value of $\langle D_0\rangle$, because we
only need the worst-case result of the final key rate, rather than
the worst-case result in each steps.

According to these we can  lower bound the fraction of single-photon
counts of the raw bits caused by signal pulses through equation:
\begin{equation}\label{mains}
\Delta_1'\ge \frac{\tilde{n}_{1s}'}{N_s}
\end{equation}
where $\tilde n_{1s}'$ is defined in Eq.(\ref{tns}) and the ranges
of related variables   in Eq.(\ref{tns}) can be calculated by
Eq.(\ref{main22},\ref{ndr},\ref{b0},\ref{nsr}) and {\em Fact 2}, if
conditions of Eq.(\ref{ace00},\ref{ace}) hold.

For coherent states, if the intensity is bounded by $[\mu^L,\mu^U]$
for the decoy pulses and $[\mu'^L,\mu'^U]$ for the signal pulses
then
\begin{equation}
a_k^{X}=(\mu^X)^ke^{-\mu^X}/k!,~a_k'^{X}=(\mu'^X)^ke^{-\mu'^X}/k!
\end{equation}
with $X=L,\;U$ and $k=1,2$ and
\begin{equation}
a_0^{L}=e^{-\mu^U},\;a_0^{U}=e^{-\mu^L}\;\;
;a_0'^{L}=e^{-\mu'^U},\;a_0'^{U}=e^{-\mu'^L}
\end{equation}
Note that our result is not limited to a coherent state source. To
calculate the unconditionally secure final key rate, one has to
combine   our result (eq.(\ref{mains})) with the existing theory of
finite-size QKD\cite{ILM,scar,cai}. Here we make a loose treatment
to show the effects of our Eq.(\ref{mains}) to the ket rate by
adding statistical fluctuation to $t$ and $t_1$ in Eq.(\ref{ilm}).
Detailed numerical results using the experimental data of QKD over
102.7 kilometers calculated by our formula is listed in table I. We
treat the experimental data in the following way for key rate of the
signal pulses by Eq.(\ref{ilm}): (1)$\Delta_1'$ can be calculated by
Eq.(\ref{mains}) as stated earlier. (2) Half of the experimental
data of signal pulses should be discarded due to the measurement
basis mis-match in the BB84 protocol. (3) Among the remaining half,
a quarter of them are consumed for the bit-flip test and another
quarter of them are consumed for the phase-flip test. (4) We use
$t_0=t_0(\mu')=3.580\%$ as the observed  value of both bit-flip rate
and phase-flip rate. The final key is distilled from the remaining
bits of signal pulses. The number of the remaining bits is not less
than $\tilde n_{1s}'/4$ . We use $t=t_0(\mu')$ for the bit-flip
rate, and
\begin{equation}
t_1= t_1'+ 10 \sqrt{4t_1'/\tilde n_{1s}'}.
\end{equation}
and $\tilde n_{1s}'$ is given by Eq.(\ref{tns}), $t_1'$ is
\begin{equation}
t_1'=\frac{t_0-\frac{p'a_0'^LD_0}{2 N_s'}}{\Delta_1'}.
\end{equation}
We then put all possible values of $D_0$ into Eq.(\ref{ilm})
according to {\em Fact 2} and Eq.(\ref{b0}) for the worst-case
result of key rate over $D_0$.

\begin{table}
\caption{\label{tab:table1}Secure key rate (final bits per pulse in
the unit of $10^{-6}$) vs intensity error upper bound using the
experimental data in the case of 102.7~km~\cite{peng}. The first row
lists different values of upper bounds of intensity errors,
$\delta_M$, i.e., when we want an intensity $x$, we can actually
create a pulse of any intensity in the range $[x(1 -
\delta_M),\;x(1+\delta_M)]$. Other rows list the final key rates,
which are the numbers of final bits per signal pulse after error
test, i.e., $\frac{n_f}{\tilde N_s}$ where $\tilde N_s$ is the
number of raw bits of signal pulses after error test, $n_f$ is the
number of final bits distilled from these $\tilde N_s$ raw bits. $R$
is the asymptotic key rate, $R_1$, $R_2$, $R_3$ are the
non-asymptotic key rates with the intensity of each pulses in $Y_0$
being bounded by 0, $0.5\%$, $1\%$, respectively. The intensity
fluctuations of  any decoy-pulse and signal pulses are bounded by
$\delta_M$. In the table we have given the results of key rate while
$\delta_M$ ranges from 0 to $3\%$. }
\begin{ruledtabular}
\begin{tabular}{cccccccc}
 $\delta_M$ & $3\%$ & $2.5\%$ & $2\%$ & $1.5\%$ & $1\%$ & $0.5\%$ & 0\\
 $R$ (in $10^{-6}$)  &11.03&12.09&13.15&14.19& 15.23 & 16.26 & 17.28\\
 $R_1$ (in $10^{-6}$) &1.536&2.567&3.591&4.607& 5.616 & 6.618 & 7.614\\
 $R_2$ (in $10^{-6}$) &1.506&2.537&3.561&4.577& 5.587 & 6.589 & 7.585\\
 $R_3$ (in $10^{-6}$) &1.475&2.507&3.531&4.548& 5.557 & 6.560 & 7.556\\
 \hline\end{tabular}
\end{ruledtabular}
\end{table}
The detailed experimental parameters can be founded in
Ref.\cite{peng}. For completeness, we list the main parameters here
in table II:
\begin{table}
\caption{\label{tab:table2}Main parameters and observed results in
the experiment of Ref\cite{peng}. $M:$ total pulses sent out by
Alice during the experimental time. $t_0(\mu),$ $t_0(\mu')$: quantum
bit error rates (QBER) of decoy pulses and signal pulses. $S', S,
S_{0}$: counting rates  of the signal pulses, decoy pulses and $Y_0$
pulses.  }

From the results listed in table I, we find that the effects of
source errors to the key rate is not significant (provided that the
source error is not too large.) For example, in the asymptotic case
(results in the first row), if the largest intensity fluctuation is
controlled to be less than $0.5\%$, the key rate is decreased 17.28
Hz(the data in the last column of the first row )  to 16.26 Hz
only(the data in the second column  from the right of the first
row). However, the statistical fluctuation decrease the key rate
significantly, given the existing experiment\cite{peng}. If we
strictly considered the finite-size effect\cite{ILM,scar,cai}, the
key rate would be further decreased. To reduce the effects, one can
increase the number of pulses in the set-up.
\begin{ruledtabular}
\begin{tabular}{ccccccccc}
 $M$ & $t_0(\mu')$ & $t_0(\mu)$ & $S'$ & $S$ & $S_0$ & $p'$ & $p$ & $p_0$\\
 $5.222\times 10^9$  & $3.580\%$ & $9.098\%$ & $1.262\times 10^{-4}$ & $4.611\times 10^{-5}$
 & $6.711\times 10^{-6}$ & 0.5 & 0.4 & 0.1\\
 \hline\end{tabular}
\end{ruledtabular}
\end{table}
\section{Some remarks on the source errors}
By the existing technology, one can indeed use whatever stabilizer
to reduce the source errors considerably. Even with this fact, our
theoretical results are still necessary: First, no technology can
guarantee a perfectly stable source. It is also questionable whether
any existing device has been proven to be able to reduce the source
errors to a low level which is exponentially close to 0. If {\em
unconditional security} is our goal in practice, we must show {\em
quantitatively} the effects of any polynomially imperfections.
 For example, even
we can control the intensity fluctuation to be less than $0.5\%$, we
had better still consider the effects of intensity fluctuation {\em
quantitatively} rather than simply {\em trust} that the effects are
negligible by {\em intuition}. Otherwise, one may ask why the cut
off point is $0.5\%$ rather than $0.0001\%$ or $25\%$ ? The security
is then  standardless.  If we disregard the effect of source errors,
possibly we shall encounter the following ridiculous story: Set-up A
produces a key rate of 1k/s with source errors of less than $0.5\%$,
set-up B produces a key rate of 20k/s with source errors of $25\%$
over the same distance. Do we have to believe that set-up B is
better ? With our theoretical results, such type of issue is
immediately resolved because we can calculate the net final key rate
after considering the effects of source errors. Second, with our
theoretical results, we don't have to blindly take too much costs in
improving the source quality. For example, with our results we now
know with strict proof that source intensity fluctuation less than
$0.0001\%$ is not so necessary, since it only improves the key rate
negligibly, if we can already control the errors less than $0.5\%$.
Finally, there are cases Eve. can indeed know the intensity
errors\cite{tdecoy}. As explicitly shown in\cite{rep,tdecoy}, in
such cases Eve. can then violates Eqs.(\ref{0011cc}) through
producing a time-dependent channel transmittance. The situation of
the Plug-and-Play is even more serious: Eve. can actually {\em
prepare} the error of each pulse then she can violate
Eq.(\ref{0011cc}) for sure.
\section{ Concluding remark and discussions} In summary, we have
shown how to calculate the lower bound of the fraction
 of single-photon counts in the decoy-state quantum key distribution
 with both source errors
and statistical fluctuations.  By our method, all imperfections have
been taken into consideration in the largest possible errors of a
few parameters of the source states in the photon number space.
Therefore we only need to know the bound values of a few parameters
of sources instead of assuming exact values of any physical
quantity. For example,  we don't have to assume zero internal
fluctuation of of Alice's attenuation as assumed by earlier
works\cite{wangapl,yi}.

Obviously, our result here  directly applies to the so called
Plug-and-Play protocol first proposed by Gisin {\em et al.}. As
pointed out in Ref\cite{gisind,yiz}, the so called Plug-and-Play
protocol can be made secure  if the attenuation factors can be
accurately set\cite{yi,px}.
 However,  in the Plug-and-Play protocol, Eve.
 actually knows the error of
each individual pulse hence both the error-free decoy-state theory
based on Eq.(\ref{0011cc}) fails. Also, as shown in the appendix, if
the attenuation factors cannot be set accurately, the result of
Ref.\cite{yi} also fails  but our theory here works.

  Also, in a Plug-and-Play protocol, Alice receives strong pulses from Bob
and she needs to guarantee the exact intensity of the pulse sending
to Bob. It is not difficult to check the intensity, but difficult to
{\em precisely  correct} the intensity of each individual pulses.
Our theory here can help to make it easier\cite{tdecoy}: Alice
monitors each pulses.  She may either choose to do crude corrections
to the pulses or not do any corrections: She simply discards those
pulses whose intensity errors are too large (e.g., beyond 2\%), and
then use our theory to distill the final key.

It should be interesting to combine our result with the existing
theories on final key distillation in a finite size
QKD\cite{ILM,scar,cai} for the unconditional final key rate in the
finite-size decoy state QKD with an unstable source.
 \\{\bf Acknowledgement:} This work was
supported in part by the National Basic Research Program of China
grant No. 2007CB907900 and 2007CB807901, NSFC grant No. 60725416,
60525201 and 60708023, and China Hi-Tech program grant No.
2006AA01Z420.
\section{Appendix}
Here we show that the elementary assumption $ Y_{m,n}^D=Y_{m,n}^S $
used in Ref.\cite{yi}is {\em incorrect} if the actively controlled
attenuation factor is not stable. (Ref.\cite{yi} assumes a stable
attenuation controlled actively.) For simplicity, we consider those
pulses containing 10 photons when flying into Alice's lab and after
Alice's attenuation, containing 1 photon when flying away from
Alice's lab to Bob. Alice decides to randomly use her transmittance
$\lambda = \lambda^D = 0.01$ to produce a decoy pulse and $\lambda =
\lambda^S=0.05$ to produce a signal pulse. However, due to whatever
un-controlled cause, there are internal fluctuations of $\lambda$
value at different times. Consider the following specific case: to
some blocks of pulses (we call these blocks {\em strong blocks}), to
all decoy pulses and signal pulses, the real transmittance is a bit
larger than $\lambda$, the value that Alice $wants$, while in some
other blocks (we call {\em weak blocks}) the real transmittance is a
bit smaller than $\lambda$. Suppose the number of pulses in the
strong blocks and that in the weak blocks are equal.  In the strong
blocks, Alice's real transmittance for a decoy pulse or a signal
pulse is $\lambda^{D+}=1.01\lambda^D$ or
$\lambda^{S+}=1.01\lambda^D$ while in the weak blocks,  Alice's real
transmittance for a decoy pulse or a signal pulse is
$\lambda^{D-}=0.99\lambda^D$ or $\lambda^{S-}=0.99\lambda^S$.
Suppose every block contains millions of pulses therefore Eve can
know whether a block is a strong block or a weak block by observe
the averaged intensity of pulses in the block. This is to say, Eve
can treat different blocks {\em differently}. Eve produces a channel
transmittance of $\eta^+$ to each pulses in the strong blocks and
another transmittance $\eta^-$ to each pulses in weak pulses. Note
that here Eve does not know which pulses are decoy pulses and which
pulses are signal pulses. She treats {\em all} pulses in one block
fixed way. Now we can directly calculate $Y_{10,1}^D$ and
$Y_{10,1}^S$:
\begin{equation}
Y_{10,1}^D =
\frac{\lambda^{D+}(1-\lambda^{D+})^9\eta^++\lambda^{D-}(1-\lambda^{D-})^9\eta^-}
{\lambda^{D+}(1-\lambda^{D+})^9+\lambda^{D-}(1-\lambda^{D-})^9}
=\frac{1.01(1-1.01\lambda^{D})^9\eta^++0.99(1-0.99\lambda^{D})^9\eta^-}
{1.01(1-1.01\lambda^{D})^9+0.99(1-0.99\lambda^{D})^9}
\end{equation}
\begin{equation}
Y_{10,1}^S =
\frac{\lambda^{S+}(1-\lambda^{S+})^9\eta^++\lambda^{S-}(1-\lambda^{S-})^9\eta^-}
{\lambda^{S+}(1-\lambda^{S+})^9+\lambda^{S-}(1-\lambda^{S-})^9}
=\frac{1.01(1-1.01\lambda^{S})^9\eta^++0.99(1-0.99\lambda^{S})^9\eta^-}
{1.01(1-1.01\lambda^{S})^9+0.99(1-0.99\lambda^{S})^9}
\end{equation}
It's easy to see that in general $Y_{10,1}^D\not= Y_{10,1}^S$ if
$\eta^+\not=\eta^-$. For example, given that $\eta^+ = 5 \eta^-$,
numerical calculation shows that
\begin{equation}
\frac{Y_{10,1}^D}{Y_{10,1}^S}>1.0025
\end{equation}
 This counter example shows that in
general the assumption $Y_{m,n}^D=Y_{m,n}^S$ is incorrect if Alice's
transmittance is not exactly controlled. We have demonstrated this
fact with a pulse containing 10 photons, one can also consider a
pulse containing $10^7$ photons and shall find very similar result:
Eve. can make $Y_{10^7,1}^D$ significantly different from
$Y_{10^7,1}^S$. This has clearly broken the elementary assumption
used in \cite{yi}.
 Obviously, such a
case is equivalent to say that the pulse intensity from Alice is
inexact, this is just the case we have studied in \cite{tdecoy}. As
was clearly shown there and also in\cite{rep}, Eve can produce
different transmittance for the single-photon state from the decoy
source and the signal source. If Alice can control the attenuation
exactly, i.e., $\lambda^D$, $\lambda^S$, there exists more efficient
way to manage the issue in the protocol of one-way quantum
communication\cite{wangapl}.


\begin{thebibliography}{99}
\bibitem{comment} As was discussed in the end of the conclusion section of
Ref.\cite{yi}:`` The security of practical QKD is a serious issue.
It is very important to implement QKD system based on tested
assumptions. There are still several crucial imperfections that are
not analyzed in this paper. For example...How can we analyze the
fluctuation of internal transmittance $\lambda$ ? How can we test
the key assumptions in our analysis, $Y_{m,n}^S=Y_{m.n}^D$ ?..."
\end{thebibliography}
\end{document}